\documentclass[12pt,manuscript]{aastex}
\usepackage{amsmath}
\usepackage{natbib}

\begin{document}

\bibliographystyle{apj}

\title{High Resolution Transmission Spectroscopy as a Diagnostic for Jovian Exoplanet Atmospheres: Constraints from Theoretical Models}

\author{Eliza M.-R. Kempton}

\affil{Department of Physics, Grinnell College, Grinnell, IA 50112}

\email{kemptone@grinnell.edu}

\author{Rosalba Perna}

\affil{Department of Physics and Astronomy, Stony Brook University, Stony Brook, NY, 11794  }
  
  \author{Kevin Heng}

\affil{University of Bern, Center for Space and Habitability, Sidlerstrasse 5, CH-3012, Bern, Switzerland}

\begin{abstract}

We present high resolution transmission spectra of giant planet
atmospheres from a coupled 3-D atmospheric dynamics and transmission
spectrum model that includes Doppler shifts which arise from winds and planetary motion.  We model jovian planets covering more than two orders
of magnitude in incident flux, corresponding to planets with 0.9 to 55
day orbital periods around solar-type stars.  The results of our 3-D
dynamical models reveal certain aspects of high resolution
transmission spectra that are not present in simple 1-D models.  We
find that the hottest planets experience strong substellar to
anti-stellar (SSAS) winds, resulting in transmission spectra with net
blue shifts of up to 3 km s$^{-1}$, whereas less irradiated planets
show almost no net Doppler shifts.  We find only minor differences
between transmission spectra for atmospheres with temperature
inversions and those without.  Compared to 1-D models, peak line
strengths are significantly reduced for the hottest atmospheres owing
to Doppler broadening from a combination of rotation (which is faster
for close-in planets under the assumption of tidal locking) and
atmospheric winds.  Finally, high resolution transmission spectra may
be useful in studying the atmospheres of exoplanets with optically
thick clouds since line cores for very strong transitions should
remain optically thick to very high altitude.  High resolution
transmission spectra are an excellent observational test for the
validity of 3-D atmospheric dynamics models, because they provide a
direct probe of wind structures and heat circulation.  Ground-based
exoplanet spectroscopy is currently on the verge of being able to
verify some of our modeling predictions, most notably the dependence
of SSAS winds on insolation.  We caution that interpretation of high
resolution transmission spectra based on 1-D atmospheric models may be
inadequate, as 3-D atmospheric motions can produce a noticeable effect
on the absorption signatures.

\end{abstract}

\keywords{planetary systems}

\section{Introduction \label{intro}}

Transmission spectroscopy has been one of the most highly utilized
methods in an astronomer's repertoire for learning about exoplanet
atmospheres.  The absorption of stellar light through a planetary
atmosphere leaves behind a spectral fingerprint of its chemical
makeup, and can provide clues as to the many physical processes
occurring in the atmosphere.  To date, measurements of exoplanet
transmission spectra have been accredited with the detection of
molecules and atoms in exoplanet atmospheres \citep[][and references
  therein]{bur09}, along with providing evidence for disequilibrium
chemistry, clouds or hazes \citep{pon13, dem13, sin13, kre14}, winds,
and orbital motion \citep{sne10}.  Typically, transmission spectra
have been obtained at moderate to low spectral resolution in order to
maximize the signal-to-noise ratio (SNR) of what is generally a very
small signature relative to the bright background of the exoplanet
host star.  More recently, the first transmission spectra have been
reported at much higher resolution ($R \approx 10^{5}$) for the planet HD 209458b \citep{sne10}.
High resolution spectra have the advantage that individual spectral
lines are fully resolved, as opposed to low resolution spectra where
only broadband features can be probed.  At high spectral resolution
and high SNR, the line profiles of individual spectral lines can be
used as valuable diagnostics of the physical processes taking place
within an exoplanet atmosphere.  This is comparable to methods used in
stellar spectroscopy, where detailed information on abundance
profiles, stellar structure, and even 3-D effects can be recovered
from high SNR spectra.  The obvious downside to high resolution
transmission spectroscopy is that many more photons must be obtained
in order to produce a high SNR spectrum.  Because of this limitation,
high resolution transmission spectroscopy has been limited so far to
the very brightest host stars harboring transiting giant planets, and
the spectra obtained still typically have quite low SNR.  This
situation is likely to improve as next generation of 30 to 40  meter class
telescopes come online.

The geometry of the transit is such that the transmission spectrum
probes a particularly interesting region of the exoplanet atmosphere.
Absorption of stellar light occurs across the terminator -- the region
of the exoplanet atmosphere that lies between the substellar (day)
side of the planet and the anti-stellar (night) side.  This region of
the atmosphere is problematic for modeling with 1-D atmospheric
models, since the transport of both heat and atmospheric gas across
the terminator can be significant.  This is especially true for
tidally-locked planets with a permanent fixed day-night boundary,
where, depending on the nature of the day-night heat transport, the
temperature difference between the two sides of the planet can be
large \citep[e.g.][]{sho09, dob10, rau10, hen11a, thr11}.  In order to
properly account for the atmospheric circulation that results, 3-D
atmosphere models are required.

Previous modeling results calculating exoplanet transmission and emission
at low spectral resolution with 3-D general circulation models (GCMs)
often find good agreement with 1-D models. However,
some inconsistencies have been shown to arise, especially in cases of strong chemical
gradients or condensation of a key absorber between the day and night
sides of the planet \citep{iro05, for10, bur10}.  These previous works have typically only considered the 3-D temperature structure of the atmosphere, while atmospheric motions have been ignored.   At high spectral resolution however, 3-D atmospheric motions cannot
be ignored, especially for cases where the magnitude of the motion exceeds the velocity resolution of the spectrograph
in use.  Observationally, this has been shown in a tantalizing manner 
by \citet{sne10}.  These authors reported a 2 $\pm$ 1
km~s$^{-1}$ blueshift in the high resolution transmission spectrum
obtained for HD 209458b.  While detected only at marginal
significance, this result has been attributed to high altitude
substellar to anti-stellar (SSAS) winds resulting from rapid flow from
the hot day-side to the cooler night-side of the tidally locked
planet.  By generating transmission spectra from 3-D atmospheric
circulation models including the Doppler shifts that arise from atmospheric motions, both \citet{mil12} and \citet{sho13} find that
day-to-night flow should produce a net blueshift of approximately 2
km~s$^{-1}$ in the transmission spectrum of HD 209458b -- in agreement
with the observational result.

The detection of Doppler shifts in exoplanet transmission spectra is
powerful indeed.  \citet{mil12} point out that high signal-to-noise
observations could be used in the future to map out the wind structure
on the approaching and receding limbs of a planet by measuring Doppler
shifts in transmission spectra during ingress and egress.  Spectral
lines originating from different locations in the atmosphere could
furthermore be used to map the wind structure as a function of
altitude.  This type of observation will likely require next
generation 30 to 40 m class telescopes to realize but could ultimately
provide a more complete picture of exoplanet meteorology.  In the
meantime, average Doppler shifts obtained over an entire portion of
the transmission spectrum can provide a measure of the mean
line-of-sight wind speed in an exoplanet atmosphere.  Previous works
by \citet{per12} and \citet{sho13} predict that tidally locked hot
Jupiters experiencing a high level of insolation should have strong
day-side to night-side winds, whereas cooler planets on longer orbits
will have less vigorous heat recirculation.  These SSAS winds are most
coherent in the upper atmosphere at pressures less than 10 mbar, which
is precisely the location in the atmosphere where transmission spectra
originate.  The effects of the SSAS wind would therefore be apparent
in the transmission spectrum via a net Doppler blue shift since the
winds are directed towards the observer while the planet is transiting
--- exactly as seen in the \citet{sne10} observations of HD 209458b.

Since the alleged detection of winds in the atmosphere of HD 209458b,
several other hot Jupiter atmospheres have been observed at high
spectral resolution using the CRIRES instrument on the VLT
\citep{kae04}, which has a resolving power of up to 10$^5$.  The hot
Jupiters $\tau$ Bo\"{o}tis b \citep{bro12, rod12}, HD 189733 b
\citep{dek13, bir13}, and 51 Peg b \citep{bro13} have had their
dayside \textit{emission} spectra observed at high resolution,
revealing molecular constituents of their atmospheres.  Furthermore, at slightly lower resolution ($R \approx 25,000$) a detection of water vapor was recently reported in the atmosphere of the non-transiting hot Jupiter $\tau$ Bo\"{o}tis b from observations using the NIRSPEC instrument  on the Keck telescopes \citep{loc14}.  While CRIRES
is currently the only instrument capable of measuring transmission
spectra at $R \ge 10^{5}$, a variety of high resolution spectrographs
are currently being planned for operation on the next generation of
30 to 40 m class telescopes.  Of these, G-CLEF on the GMT \citep{sze12},
HROS on the TMT \citep{cra08}, and the proposed CODEX \citep{pas10},
SIMPLE \citep{ori10}, and HIRES \citep{mai13}  instruments for the E-ELT will have spectral resolution of
approximately $10^{5}$, which should provide a powerful complement to
the telescopes' large mirror sizes.  Spectra obtained with these
instruments will open the door in the future to a wide array of
exoplanet atmospheric diagnostics \citep[e.g.][]{sne13}.
  
In this paper, we expand upon the previous work of \citet{mil12} and
\citet{sho13} to study the effects of 3-D atmospheric structure and
atmospheric circulation on high resolution exoplanet transmission
spectra, including the Doppler shifts that arise from winds and the planet's rotation.  We extend the previous results to the study of planets
ranging across more than two orders of magnitude in incident flux.  We
furthermore study the effects of stratospheric temperature inversions
on the spectra.  To accomplish this, we have coupled together the
transmission spectrum model described in \citet{mil09} and
\citet{mil10} with the atmospheric dynamics models of \citet{per12} to
produce transmission spectra that self-consistently treat the
underlying 3-D atmospheric structure and circulation.  The methodology
is similar to that of \citet{mil12} and \citet{sho13} but is applied
across a broader range of parameter space to study tidally locked hot
jovian planets as a population.  We furthermore extend our study to
hot Jupiters with layers of optically thick clouds, and we comment on
how high resolution transmission spectra can be used to constrain the
properties of these atmospheres.  Our modeling methods are laid out in
Section~\ref{methods}.  Results are presented in Section~\ref{results}
and we finish with concluding remarks in Section~\ref{conclusion}.

\section{Model Description}
\label{methods}

\subsection{3-D Dynamics Models \label{model}}

For our study, we use the same suite of 16 atmospheric circulation
models computed by \citet[][PHP in the following]{per12}, using the
numerical setup described in \citep{hen11b, hen11a}.  A detailed
description of the models, as well as the numerical techniques
employed, can be found in the above papers, and hence in the following
we will limit ourselves to summarize the main features and describe
the updates that have been made since the publication of PHP\footnote{
We remind that those models do not include magnetic drag, which is known
to reduce wind speeds if the atmosphere is strongly magnetized
\citep{per10a, rog14}.}.

The 16 models are computed for 8 values of the irradiation temperature
$T_{\rm irr}$ in the range $\approx 770-3000$~K.  The effective
stellar temperature $T_*$ and stellar radius $R_*$ are related to the
irradiation temperature by the relation
\begin{equation}
T_{\rm irr} = T_\star \left( \frac{R_\star}{a} \right)^{1/2},
\end{equation}
where $a$ is the semi-major axis of the orbiting exoplanet.  Given an irradiating flux ${\cal F}_0 =\sigma_{\rm SB} T^4_{\rm irr}$, this is expressed by
\begin{equation}
a = R_\star T^2_\star \left( \frac{\sigma_{\rm SB}}{{\cal F}_0} \right)^{1/2} \nonumber 
\end{equation}
\begin{equation}
\approx 0.09 \mbox{ AU}\left( \frac{R_\star}{R_\odot} \right)
\left( \frac{T_\star}{6000 \mbox{ K}} \right)^2
\left( \frac{{\cal F}_0}{2 \times 10^8 \mbox{ erg cm}^{-2} \mbox{ s}^{-1}} \right)^{-1/2},
\end{equation}
where $\sigma_{\rm SB}$ is the Stefan-Boltzmann constant and the mass
$M_*$ of the host star is taken to be one solar mass.  The planet is
assumed to be tidally locked, which is generally a good approximation
for the models considered here (see PHP for details).  The orbital
frequency (equal to the rotational one), is then given by $\Omega_p
\approx \left({G M_\star}/{a^{3}}\right)^{1/2}$.  With this setup, for
each irradiation temperature $T_{\rm irr}$, the values for ${\cal
  F}_0$, $\Omega_p$ and $a$ are correspondingly determined according
to the relations above.  The numerical values of these quantities are
reported in Table~\ref{tab:params}.

%For each temperature value, a case with and without temperature inversion is considered.  This is accomplished by computing models with two different values for $\gamma$, the ratio between the opacity for incoming radiation to that of the outgoing radiation.  The two values considered are $\gamma = 0.5$ and $\gamma = 2.0$, corresponding to non-inverted and inverted temperature profiles, respectively.  

For each of the 8 values of the irradiation temperature in
Table~\ref{tab:params}, we consider a circulation model with
temperature inversion, and another without.  In the absence of clouds
and hazes, the use of the dual-band approximation for radiative
transfer \citep[e.g.][]{hen11a} allows for establishing the presence
of a temperature inversion by means of the magnitude of the ratio
$\gamma_0 \equiv \kappa_{\rm S}/\kappa_0$, where $\kappa_{\rm S}$ is
the shortwave (optical) opacity, and $\kappa_0$ is the longwave
opacity.  If $\gamma_0 > 1$, the shortwave photosphere sits above the
longwave photosphere and an inversion exists; if $\gamma_0 < 1$, the
relative locations of the two photospheres are reversed, and there is
no inversion. For the former situation, we consider a case with
$\gamma_0=2$, while for the latter we take $\gamma_0=0.5$.  As described in PHP, the simulations make the simple assumption that temperature inversions (or their absence) may be maintained over the entire atmosphere, as a means of isolating their effects on the results, commensurate with the fact that the identity of these "stratospheric absorbers" remains basically unknown.  

\begin{figure}
\begin{center}
\includegraphics[scale=0.90]{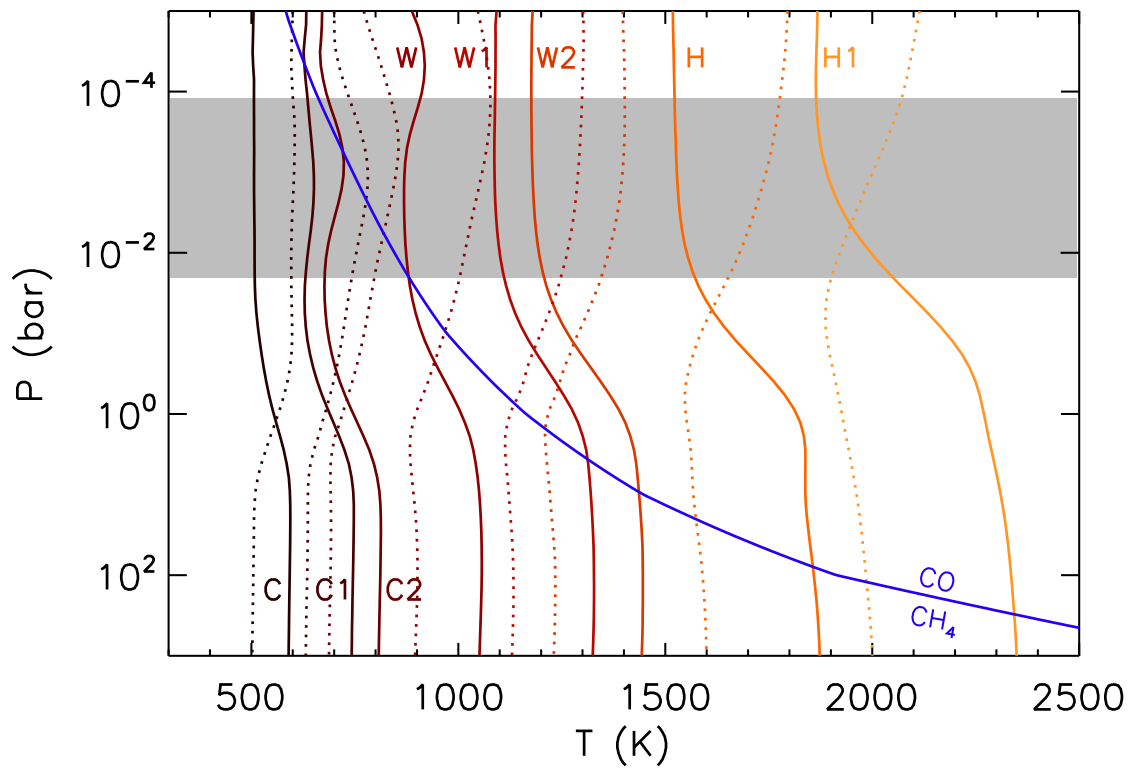}
\end{center}
\caption{Global mean T-P profiles for the models presented in this
  paper.  The profiles shown were computed by taking a planet-wide average of the individual T-P profiles from our model grid of 96 latitude by 192 longitude grid points.  The gray shaded region indicates the portion of the atmosphere typically probed by transmission spectroscopy.  The CO /
  CH$_4$ equilibrium line is indicated in blue. \label{fig:t_p_avg}}
\end{figure}

Global-mean temperature-pressure profiles for all 16 models are shown
in Fig.~\ref{fig:t_p_avg}, with the pressure range relevant for
absorption studies (approximately 0.1 - 10 mbar) indicated.  Since the
publication of PHP, the upper boundary of the models has been extended
to 10 $\mu$bar to include the portion of the atmosphere probed by
transmission spectroscopy.  The
equilibrium condition for CO and CH$_4$ to be present in equal
quantities is indicated in Figure~\ref{fig:t_p_avg} as well.  For all
of our models except the coldest ones (C, C1, and C2) CO is expected
to be the dominant carbon-bearing molecule in the region of the
atmosphere probed by transmission spectroscopy.  It should be noted
however that 3-D effects can produce more (or less) CO in hotter (or
colder) regions of the atmosphere, and non-equilibrium chemistry may
further alter the abundances of carbon-bearing molecules beyond their
calculated values \citep{coo06}.  The former effect is accounted for
in our modeling.  The latter is not, as we only compute abundances in
thermochemical equilibrium \citep[see][]{mil09}

\subsection{Transmission Spectroscopy Model}

We have coupled the transmission spectroscopy model described in
\citet{mil09} and \citet{mil10}, with the 3-D atmospheric circulation
models from PHP.  The coupling of the two models is essentially the
same as the one described in \citet{mil12}.  Briefly, light rays are
passed through the upper atmosphere of the planet, accounting for the
geometry of the grazing trajectory for stellar light through the
planetary atmosphere during transit.  The attenuation of stellar light
is calculated by integrating the transfer equation for the case of
absorption only, along the path of the light
\begin{equation}
I(\lambda) = I_{0} e^{-\tau}, 
\label{intensity}
\end{equation}
where $I_0$ is the incident intensity from the star.  The total
absorption is then found by integrating the individual light rays over
the entire annulus of the planetary atmosphere, as viewed by an
observer on Earth.
 
At each location in the atmosphere --- defined by its latitude,
longitude, and height coordinates --- the opacity, $\kappa$ (in units of cm$^2$ g$^{-1}$), is
determined for solar composition gas in chemical equilibrium at the
local temperature and pressure.  The optical depth $\tau$ from
Equation~\ref{intensity} is then obtained according to
\begin{equation}
\tau = \int \rho \kappa ds,
\label{tau}
\end{equation}
where $ds$ is the differential path length computed along the
observer's line of sight, and $\rho$ is the gas density.  In order to account for atmospheric motions
resulting from winds and the planet's rotation, we Doppler-shift the
opacity at each location in the atmosphere according to the local
line-of-sight (LOS) velocity.  At the center of transit time, the LOS
velocity is given by
\begin{equation}
v_{LOS} = -[u \sin \theta \cos \phi + v \cos \theta \sin \phi + (R_{p} + z) \Omega_{p} \sin \theta \cos \phi]
\label{vlos}
\end{equation}
where $u$ and $v$ are the zonal and meridional components of the
windspeed respectively, $\phi$ and $\theta$ are the latitude and
longitude, $R_{p}$ is the planetary radius, and $z$ is the altitude
above the 1-bar pressure location.  We neglect the vertical component
of the windspeed, since it does not contribute significantly to
$v_{LOS}$ at the terminator due to geometric effects.  Furthermore,
the vertical winds are typically at least two orders of magnitude
smaller than the zonal ones.  By incorporating the Doppler-shifted
opacities into our 3-D calculation, we produce transmission spectra
that are consistent with the underlying 3-D atmospheric structure and
wind pattern.

\section{Results}
\label{results}

\subsection{Temperature-Dependent Effects}

Spectra for each of the 16 models are shown in
Figure~\ref{fig:spec_unshifted} for a calculation without the Doppler
shifting effects of winds or planetary rotation included.  The spectra
appear qualitatively similar for the models with and without
temperature inversions.  The strong resemblance between models with
very different temperature structure is unsurprising.  Transmission
spectra arise from pure absorption through the atmosphere, which has
only a minor dependence on temperature, insofar as the gas opacities
(and associated chemistry) vary slowly as a function of $T$.  For the
same reason we also find that the spectra calculated without any
Doppler shifts applied are qualitatively similar to spectra calculated
using a 1-D T-P profile that is the average of all of the profiles
from the 3-D model (not shown), in agreement with prior work by
\citet{for10}.  Going from hottest to coldest, the most obvious change
in the unshifted spectra is the decrease in line strength that results
from the reduction in atmospheric scale height, which is in turn
proportional to temperature.  Also, for the very coldest models, more
carbon is incorporated into CH$_4$ than CO, resulting in a complex
forest of CH$_4$ lines appearing in the spectra for these models.  The
transition from CO to CH$_4$ takes place at lower $T_{irr}$ for the
models with temperature inversions than for those without.  Even for
the very coldest model (C), some CO absorption still remains in the
transmission spectrum for the model with $\gamma =2.0$.  At all levels
of planetary irradiation, the CO absorption is slightly stronger in
the temperature inversion models, owing to the higher temperature at
altitude, which produces a correspondingly larger scale height along
with slightly higher CO abundance in the region where the transmission
spectrum originates.  This is a very small effect however, and the
unshifted spectra imply that it will remain very difficult to
distinguish between models with differing temperature structure using
transmission spectra observations.

When the Doppler shifts resulting from winds and planetary rotation
are self-consistently included in the modeling of the transmission
spectra, peak line strengths are considerably reduced and individual
spectral lines are correspondingly broadened (see
Figure~\ref{fig:spec_winds}).  The spectra for the hottest planets are
substantially blue shifted, which we discuss in detail in the
following subsection.  For tidally locked planets, the hottest models
experience the most significant rotational broadening.  This results
from the short orbital periods and correspondingly short rotation
periods for the hottest planets.  As a result, the larger scale height
of the hot atmospheres, which would typically produce much stronger
absorption features, is mitigated to a large extent by a reduction in
line strength from more extreme rotational broadening.  The advantage
of observing hot planets in transmission is therefore substantially
reduced. For high resolution spectra under conditions where the spectral resolution $R > c /
v_{rot}$, it should be noted that 1-D models that do not at least
include effects of rotational broadening might over-predict the
strength of spectral lines in transmission spectra by a factor of
several.

\subsection{Atmospheric Winds}

In addition to the rotational broadening, atmospheric circulation for
the hottest jovian planets has an appreciable effect on the
transmission spectra.  Whereas rotational broadening has an (almost)
symmetrical effect on spectral lines, winds can produce highly
asymmetric perturbations to spectral lines profiles
\citep[e.g.][]{mil12, sho13}.  The effects of winds can only be
modeled through 3-D calculations such as the ones presented here.  1-D
models entirely miss out on the significant effects of winds on
transmission spectra, which become apparent at high resolution, as
seen in Figure~\ref{fig:spec_winds}.  The transmission spectrum
originates from high in the atmosphere ($P_{atm} \sim$ 1 mbar), where
the atmospheric circulation is predominantly advective, resulting in
direct transport of hot dayside gas across the terminator to the
colder night side of the planet.  The transit geometry guarantees that
the day-to-night winds will be directed toward the observer along
his/her sightline, producing a net blue shift in the absorption
signature of the exoplanet atmosphere.

The blue shifting of hot Jupiter transmission spectra is apparent from
our models in Figure~\ref{fig:spec_winds}, where a slight shift to
shorter wavelengths can be seen by eye in the spectra of the hottest
planets.  Wind speeds are typically larger for the hotter exoplanets,
and the high altitude day-to-night flow is more coherent, resulting in larger net blueshifts for the
more highly irradiated atmospheres.  The strong SSAS flow for the
hottest models can be explained in the following way, as pointed out
originally by \citet{sho02}.  In highly irradiated gas giants, the
flow is expected to occur roughly at the sound speed.  This means that
the dynamical timescale can be given by $t_{dyn} \sim R_p/c_{s}$.  It
follows that
\begin{equation}
\frac{t_{rad}}{t_{dyn}} \sim \frac{c_{P} P c_{s}}{g \sigma_{SB} T^{3} R_{p}} \propto \frac{P}{T^{5/2}}. \label{eq:timescale}
\end{equation}
The gas heat capacity, surface gravity, and planetary radius ($c_P$,
$g$, and $R_{p}$) are held constant in our simulations, and the
relevant pressure is taken to be the photospheric pressure.
Atmospheres with higher values of $t_{rad}/t_{dyn}$ are therefore
colder and have flows that are more zonal in nature with comparable
red- and blue shifted components.  Those with lower values of the
ratio are hotter and have more prevalent SSAS flow.

The LOS components of the exoplanet winds around the terminator are shown in Figure~\ref{paul_figure}.  The behavior is essentially as described above.  The winds become significantly weaker for models with lower $T_{irr}$.  Additionally, the suite of models with the exception of the coldest one (model C) displays a slow transition from SSAS flow to more zonal flow as the irradiating flux is diminished.  In the case of model C, the flow transitions abruptly back to being strongly SSAS, which likely results from differences in opacities.  Model C is the only one in the suite where carbon is almost exclusively bound up in CH$_4$, rather than CO, which affects the radiative transfer.  

\begin{figure}
\begin{center}
\includegraphics[scale=0.61]{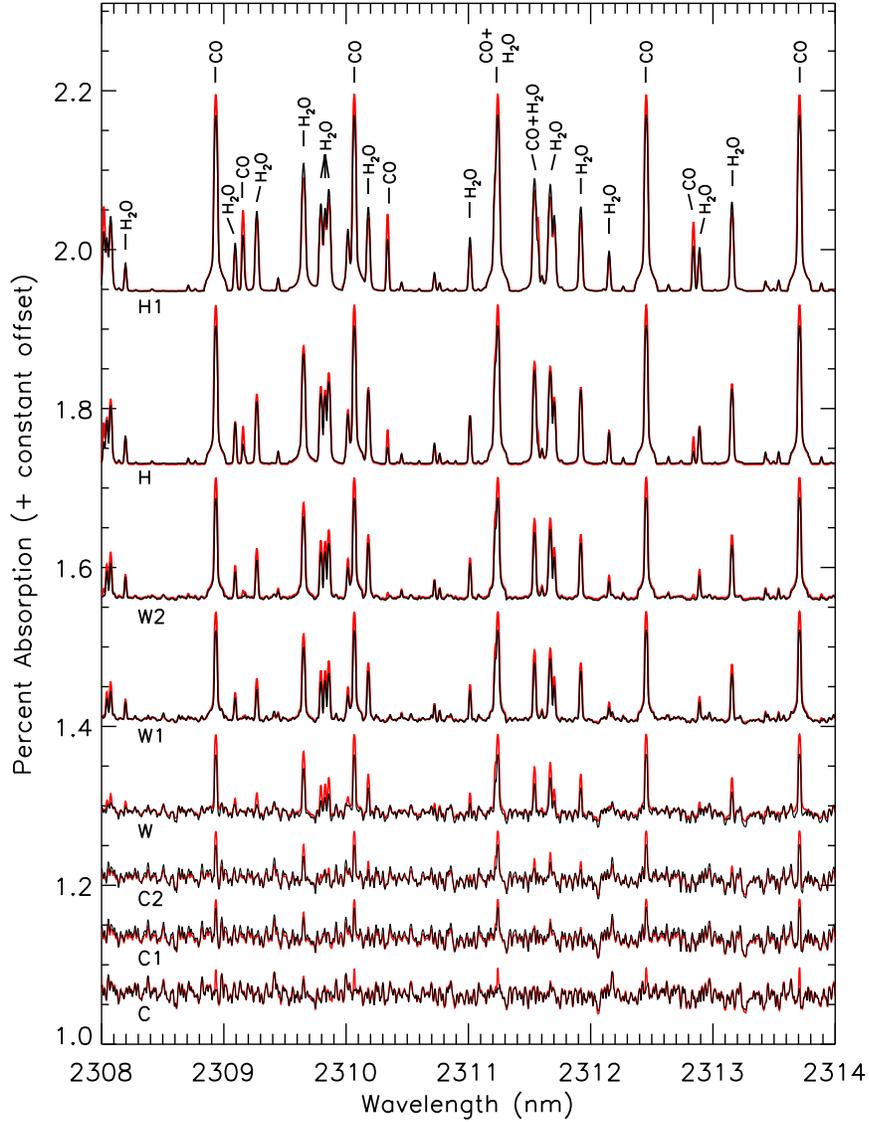}
\end{center}
\caption{Transmission spectra for the 16 model atmospheres considered
  in this paper.  No Doppler shifting of spectral lines has been
  applied to these spectra.  Models go from hottest at the top, to
  coldest at the bottom.  Red lines are for atmospheres with
  T-inversions.  Black lines are atmospheres with no inversions.  An
  arbitrary vertical offset has been added to each spectrum for easier
  viewing.  CO and H$_{2}$O absorption features are indicated.  (The
  vast majority of the unmarked features in the spectra of the colder
  planets are due to methane.  The flat "continuum" opacity source in the hottest models results from collision-induced absorption of H$_2$-H$_2$ and H$_2$-He pairs.)} \label{fig:spec_unshifted}
\end{figure}

\begin{figure}
\begin{center}
\includegraphics[scale=0.61, angle=90.0]{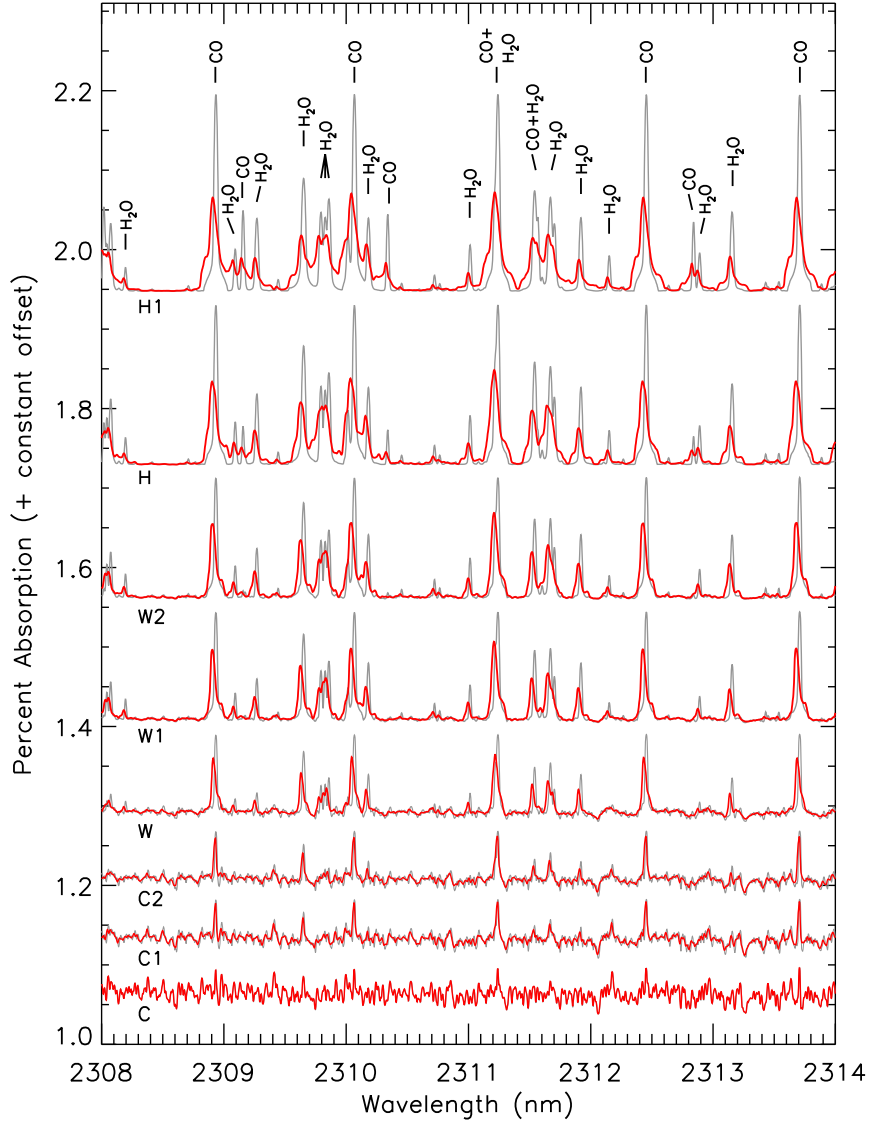}
\end{center}
\caption{ Doppler shifted transmission spectra for the 8 models with
  temperature inversions ($\gamma = 2.0$, red lines).  Gray lines are
  the unshifted spectra from Figure~\ref{fig:spec_unshifted}, shown
  for reference.  An analogous figure for the non-inverted models
  ($\gamma = 0.5$) is not shown but looks qualitatively the same.
  Rotational broadening and a net blueshift due to day-to-night winds
  is most apparent for the hottest models.} \label{fig:spec_winds}
\end{figure}

At this point in time, observers are typically not able to obtain high
resolution transmission spectra at high signal-to-noise, like the ones
shown in Figures~\ref{fig:spec_unshifted} and~\ref{fig:spec_winds}.
Instead, information about the transmission spectrum is recovered by
cross correlating the entire observed spectrum against a template.  The template is typically chosen to be a modeled spectrum of the system in question.
The best fit parameters of the planetary atmosphere are then inferred
by locating the highest significance peak in the cross correlation
function.  For consistency with the observational method, we have
cross-correlated our Doppler-shifted transmission spectrum models
against their unshifted counterparts (shown in Figure~\ref{fig:spec_unshifted}), which serve as template spectra.
The net Doppler shift obtained from the cross correlation process
gives a measure of the average LOS velocity in the upper atmosphere of
the planet that would be recovered observationally.

\begin{figure}
\begin{center}
\includegraphics[scale=0.92]{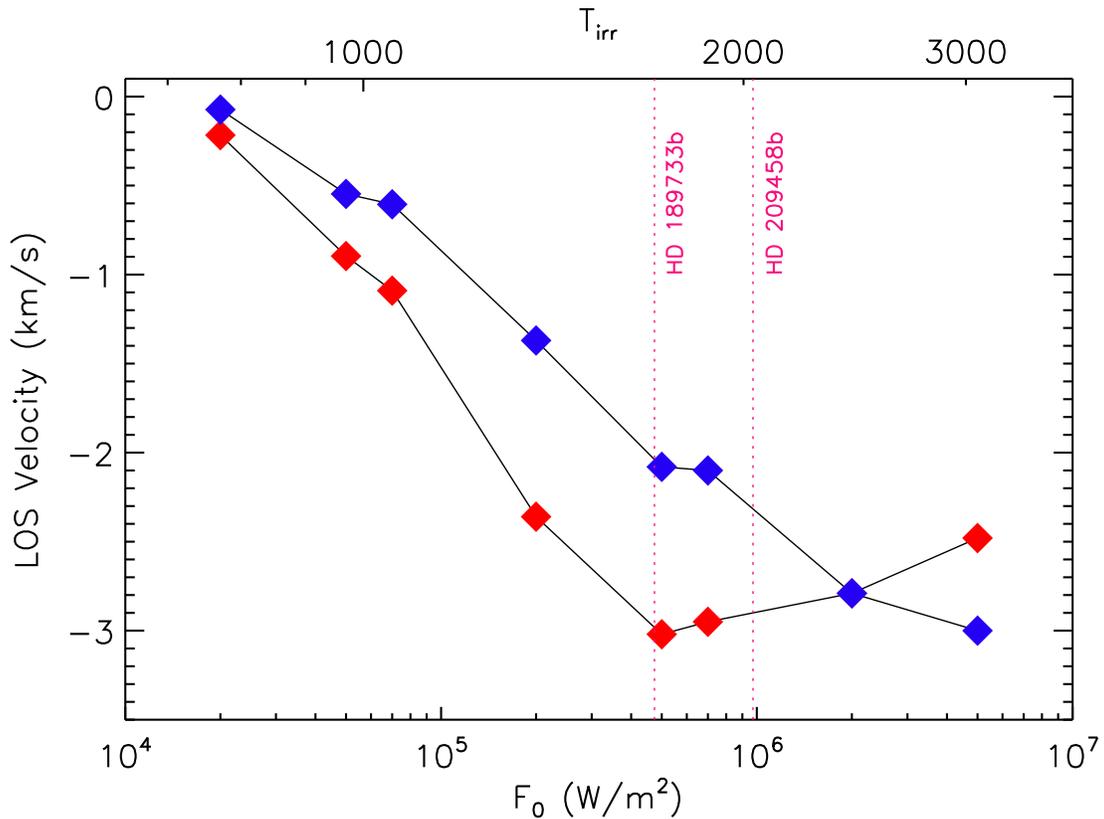}
\end{center}
\caption{LOS velocities for each of the 16 models studied in this
  paper, obtained by cross-correlating the fully Doppler shifted
  transmission spectra against an unshifted template spectrum.  Models
  with temperature inversions ($\gamma = 2.0$) are shown in red.
  Those without T-inversions ($\gamma = 0.5$) are shown in blue.  The
  irradiation levels for the hot Jupiters HD 209458b and HD 189733b
  are indicated.} \label{fig:doppler_shifts}
\end{figure}

Our results from the cross correlation process can be seen in
Figure~\ref{fig:doppler_shifts}.  We find large net blue shifts of 2 -
3 km s$^{-1}$ for the hottest exoplanets, and there is a gradual shift
to no discernible blue-shift for the least irradiated of our models.
Planets both with and without temperature inversions show the same
qualitative behavior.  However, at most levels of irradiation, the
models with temperature inversions produce larger net blue shifts by
up to about 1 km s$^{-1}$.  This follows again from
Equation~\ref{eq:timescale} --- the transmission spectrum probes
hotter regions of the atmospheres with inversions, which are more
radiative, thus producing stronger SSAS flow.  As a result, the
gradual transition from large blue shift to no blue shift begins at a
lower irradiation level for the T-inversion models than for those
without inversions.  For the $\gamma = 2.0$ models, we predict that HD
209458b and HD 189733b should both have large net Doppler blue shifts
in their transmission spectra of $\approx 3$ km s$^{-1}$, whereas the
$\gamma = 0.5$ models predict a weaker blue shift for HD 189733b than
for HD 209458b.  These conclusions are in line with the work of
\citet{sho13} who predicted a smaller net blue shift for the less
irradiated HD 189733b, although with a larger rotational broadening
component due to HD 189733b traversing a tighter orbit around a cooler
star than HD 209458.  Observational confirmation of the net blue
shifts for the transmission spectra of these two planets in
particular, along with other giant planets orbiting bright stars, may
be possible in the coming years, as evidenced by the work on HD
209458b by \citet{sne10}, which would provide useful validation of our
modeling results.

\subsection{Clouds}

Recently, convincing evidence has been presented that clouds and/or
hazes play a considerable role in the transmission spectra of a wide
variety of exoplanet atmospheres \citep[e.g.][]{kre14, sin13, pon13}.
The nature of the cloud particles is unknown at this time, but they
appear to be a prevalent feature in irradiated exoplanet atmospheres,
and many radiative transfer models are poorly equipped to deal with
their presence.  From the observational standpoint, clouds are a
hindrance to inferring the composition of an exoplanet atmosphere via
transmission spectroscopy.  Layers of high altitude clouds or hazes
can create an optically thick barrier, preventing observation of
deeper layers of the atmosphere where atomic and molecular spectral
features would otherwise originate.  In low spectral resolution
observations, this effect can also be misconstrued with the reduced
scale height associated with high metallicity.  The flattening or
smoothing of the transmission spectrum as a result of clouds is a
well-documented phenomenon that can confound efforts to measure the
composition of an exoplanet's atmosphere.  This has been particularly
problematic in the case of the super-Earth GJ 1214b.  Measurements of
the bulk density of the planet imply that it must have a substantial
gaseous atmosphere \citep{rog10, net11, val13}.  However, observations
of the planet's transmission spectrum at ever increasing precision
continue to reveal a flat spectrum, which is now only consistent with
the presence of a high thick layer of clouds \citep{kre14}.  To
resolve the problem of what such an atmosphere is composed of, high
resolution spectroscopy may be the answer.  While at low resolution,
the transmission spectrum might appear entirely flat, narrow strong
spectral lines should still remain optically thick above the cloud
deck.  At high spectral resolution, these lines could therefore be
observed, and an atmospheric composition inferred from their presence.

\begin{figure}
\begin{center}
\includegraphics[scale=0.8, angle=90.0]{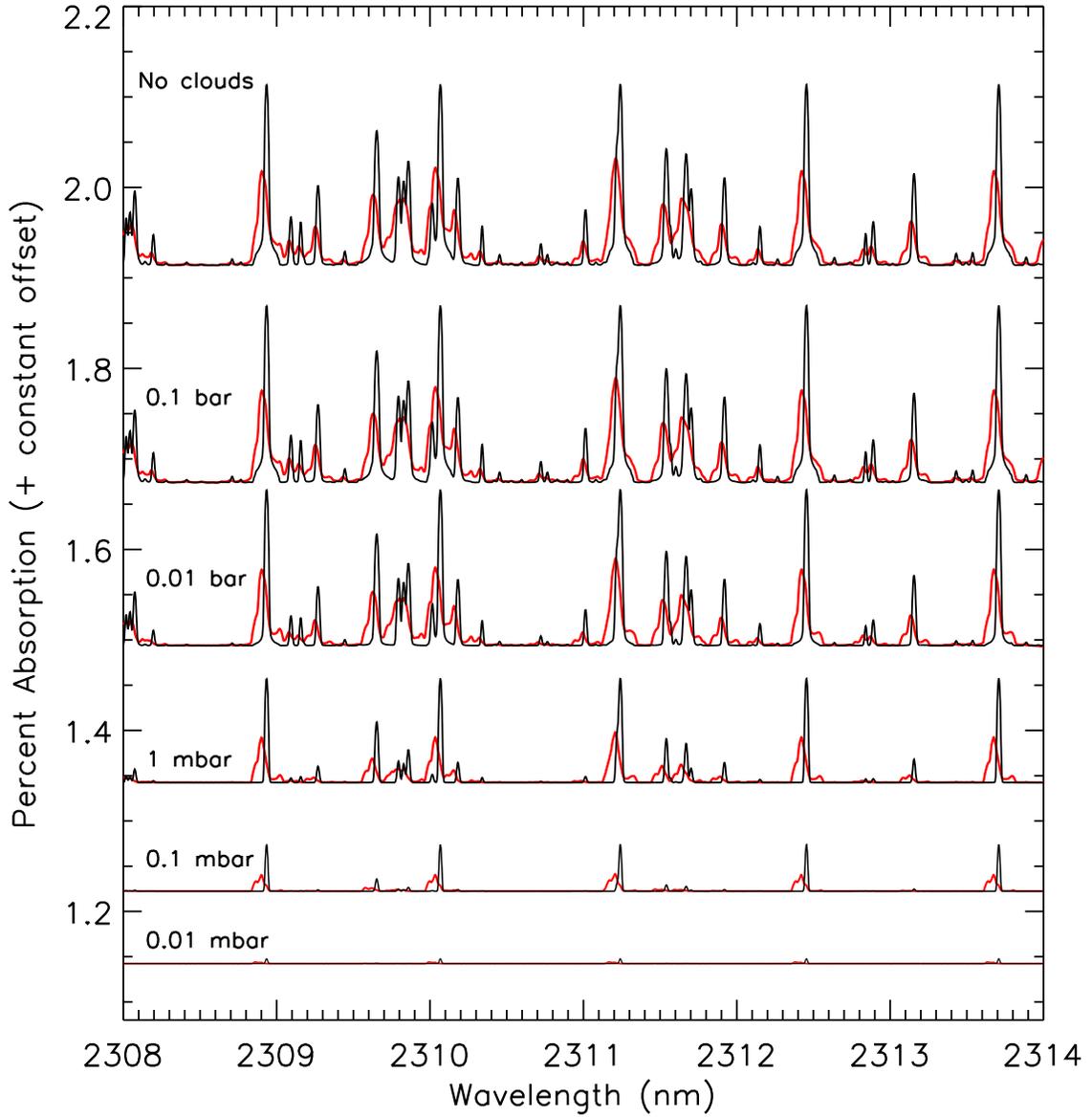}
\end{center}
\caption{Transmission spectra for the second hottest model of our
  suite (Model H) with $\gamma = 2.0$.  Optically thick clouds have
  been inserted at pressures from 0.1 bar to 0.01 mbar.  Spectra
  without Doppler shifts introduced are in black, and those with
  Doppler shifts caused by clouds and planetary rotation are in red.
  The reduction in line strength for clouds at higher altitude results
  from stellar light permeating a much reduced portion of the
  atmosphere.} \label{fig:clouds}
\end{figure}

We have performed limited tests of the effects of high altitude clouds
on hot Jupiter transmission spectra by truncating absorption of
stellar light at the height of an arbitrarily inserted cloud deck.
Clouds are inserted at 100, 10, 1, 10$^{-1}$ mbar, and 10$^{-2}$ mbar,
and the results for one of our model atmospheres (model H) are shown
in Figure~\ref{fig:clouds}.  The signal drops precipitously as the
height of the cloud deck is increased, however some signal remains
present up to cloud decks at 0.1 mbar.  While it will likely prove
challenging observationally to detect molecular signatures from above
a high cloud deck, high resolution transmission spectroscopy of
narrow, but strong, spectral features may provide the only tenable
path forward in this regard.

\section{Conclusions}
\label{conclusion}

Our grid of jovian planet atmosphere models reveals that transmission
spectra at high resolution can be used to probe a variety of
atmospheric physics.  We have found that even at very high spectral
resolution, transmission spectra are not strongly dependent on the
temperature structure for an atmosphere.  While it is unfortunate that
transmission spectra do not reveal much information about atmospheric
temperature structure, this removes a significant source of degeneracy that is
typically present in the interpretation of emission spectra.  As a
result, transmission spectra can be used as a relatively unambiguous
probe for atmospheric composition and wind structure, without the
worry of convoluting these effects with other temperature-dependent
effects.  While we find some minor differences between transmission
spectra for atmospheres with temperature inversions relative to those
without, these differences are likely too small to be observable in
the near future.

The effects of winds on jovian planet transmission spectra however,
can be substantial.  Significant rotational broadening is expected for
hot Jupiters on close-in orbits, and day-to-night winds will imprint a
net blueshift on the transmission spectra of up to several
km~s$^{-1}$.  The finding that the most strongly irradiated jovian
planets should produce transmission spectra with the largest net
Doppler blue shifts provides a testable hypothesis from our results.
Previous studies that claim close resemblance between 1-D and 3-D
modeled transmission spectra \citep{for10} remain valid at low
spectral resolution, but we have shown here that 3-D models depart
significantly from their 1-D counterparts when dynamical effects are
taken into account, especially for the hottest exoplanets.

\begin{figure}
\begin{center}
\includegraphics[scale=0.76]{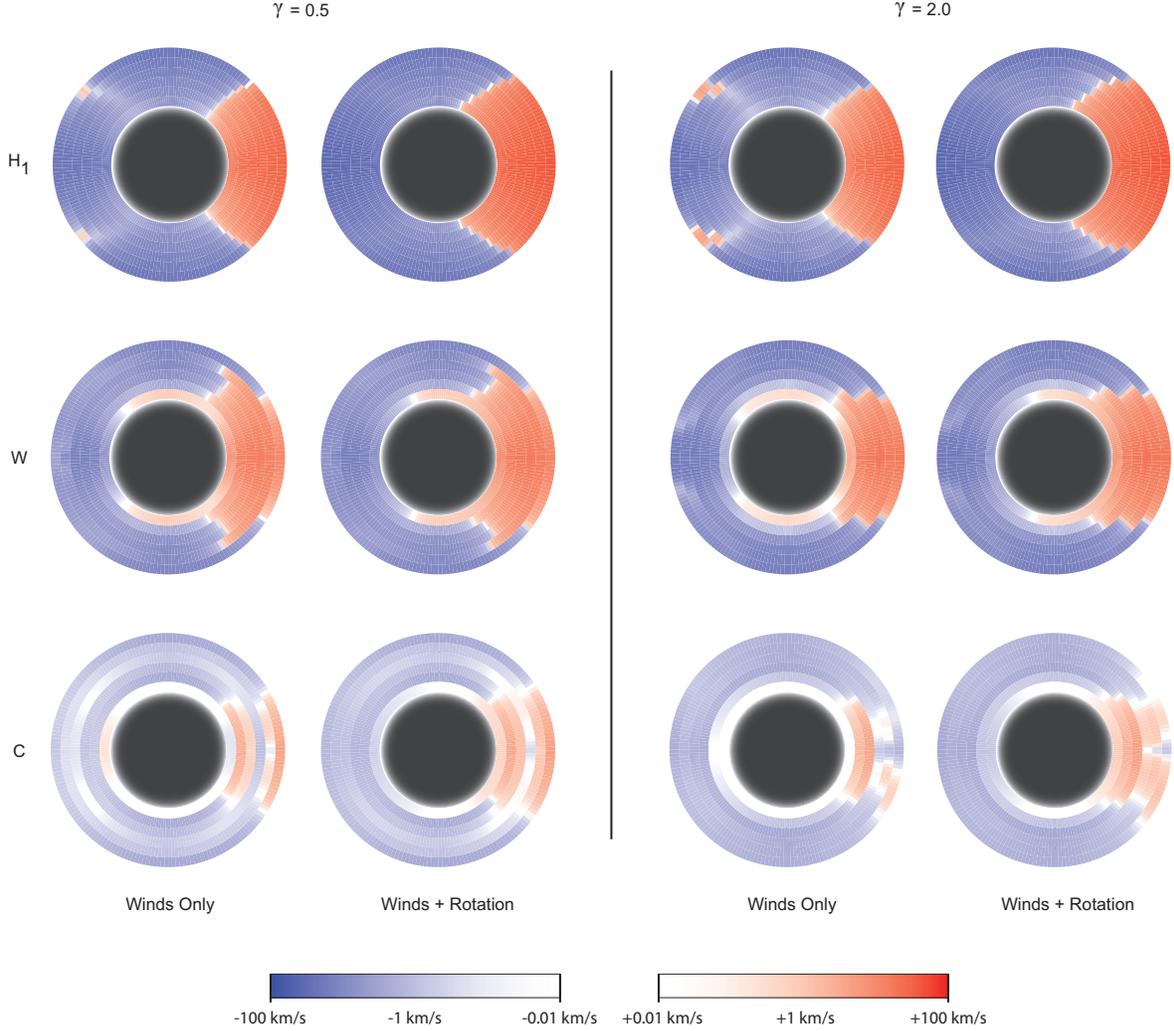}
\end{center}
\caption{Projected average LOS velocities along the observer's sight
  line through the atmospheres of models C, W, and H1.  Darker blue
  (red) corresponds to stronger blue (red) shifted winds.  For each
  pair of images, the leftmost includes only the effects of winds,
  while the rightmost includes the effects of both winds and planetary
  rotation.  The hottest models have the highest wind speeds and the
  fastest rotation.  Max LOS wind speeds for the C, W, and H1 models respectively are 11, 4, and 0.5 km s$^{-1}$ (18, 5, and 0.6 km s$^{-1}$ when rotation is included in the calculation).  LOS velocities have been calculated at each of 6
  pressure levels -- 1 bar, 0.1 bar, 0.01 bar, 1 mbar, 0.1 mbar, and
  0.01 mbar.  Pressure increases from the outermost annulus, inward to
  the dark circle at the center of each image, which represents the
  optically thick interior of the planet.  This figure is not to scale.} \label{paul_figure}
\end{figure}

Our work in this paper extends the previous studies of \citet{sho13} who looked at three specific exoplanets -- HD 209458b, HD 189733b, and GJ 436b -- to see how stellar insolation affects the atmospheric dynamics and alters the resulting Doppler shifted transmission spectra.  The authors reported a "regime shift" between strong SSAS flow for the heavily irradiated HD 209458b and zonal flow for the colder planet GJ 436b.  In our current work, rather than modeling three individual planets with different radii, surface gravities, and stellar hosts, we provide a well-defined (and well-sampled) grid of models to more carefully study the effects of insolation on Doppler shifted transmission spectra.  In our suite of models we observe a similar shift in upper atmosphere dynamics from SSAS flow for hot planets to zonal flow colder planets.  However, our shift is not as pronounced as what was reported in \citet{sho13}, and we would hesitate to call it a regime shift.  In fact, the development of jets and a zonal wind pattern appears to reverse itself in our very coldest models.  Our coldest model (model C) has a strongly SSAS flow (see Figure~\ref{paul_figure}), which likely results from changes in atmospheric carbon chemistry, which in turn affects opacities and the resulting radiative transfer at low levels of insolation.  Like \citet{sho13} we do however, continue to see a reduction in the net Doppler blueshift as the level of insolation is reduced.  This trend continues, all the way to the very coldest models.  Our model C produces the smallest net blueshift overall, despite having strong SSAS flow in the upper atmosphere.  The reason for this is straightforward -- the average wind speeds fall off sharply with decreasing $T_{irr}$.  Despite having coherent SSAS flow in model C, the average wind speeds are so small that there is very little discernible blue shift in this model.  As a result, we conclude that Doppler shifted transmission spectra will only be useful in determining the extent of SSAS vs. zonal flow for the most highly irradiated jovian planets.  Cooler giant planets will have slower wind speeds overall, and it therefore becomes much more challenging to use Doppler-shifted transmission spectra to diagnose the planet-wide wind structure -- even when observations are obtained with very high spectral resolution.  

As for why we do not observe a strong regime shift in our models as a function of insolation, the differences between our suite of models and those of \citet{sho13} are difficult to fully diagnose without performing a true one-to-one comparison.  However, we can offer up a couple of suggestions.  First of all, the treatment of atmospheric drag is not identical between our two suites of models.  It has previously been shown in both \citet{mil12} and \citet{sho13} that the treatment of atmospheric drag can have substantial effects on atmospheric circulation, which in turn alters the degree of blue-shifting of the transmission spectra.  In our suite of models we do not provide any prescription for drag from magnetic fields or shocks, but our simulations do explicitly include numerical dissipation to stabilize against numerical noise such that they are able to run to completion.  Generally, simulations of atmospheric circulation inevitably include some form of numerical dissipation, even if not explicitly applied, via the computational grid and numerical scheme.   Our choice, as laid out in PHP, has been to apply a single planet-wide numerical dissipation factor that is proportional to $1/\Omega$, where $\Omega$ is the rotational frequency.  \citet{hen11a} point out that wind speeds from 3-D GCMs are uncertain at the $\sim$ 10\% level, depending on how one specifies the numerical dissipation / drag.  This is likely to be the key source for the differences in circulation patterns between our own models and those of \citet{sho13}.  Another  potential difference comes from the rotation rates of our models.  All three planets modeled in \citet{sho13} have similar orbital (and therefore rotational) periods of close to 3 days, whereas our coldest modeled planets have rotation periods of approximately 56 days.  These differences in rotation rate may bring about subtle changes to the atmospheric dynamics. \citet{sho09}, \citet{kat13}, and \citet{rau14} have studied the effects of rotation rate on atmospheric dynamics for non-tidally locked planets and/or planets on eccentric orbits, and all of these authors report substantial changes in atmospheric circulation by altering the planetary rotation rate.  However, it is not clear how relevant these previous results are to our current conclusions, as the dynamical effects produced in the three aforementioned papers probably result in large part from the very different stellar forcing brought about by the non-tidally locked and eccentric orbital configuration of their models.

The use of cross correlation techniques to recover information from
high resolution transmission spectra results in the entire spectrum
being used simultaneously for interpretation, rather than direct
observations of individual spectral features.  These techniques
typically rely upon a model spectrum as a template for the spectral
recovery process, and the template spectra have typically have been
produced from 1-D radiative transfer models.  The assumption is that
the template is an accurate representation of the observed spectrum.
If this is not the case, inconsistencies may arise in the
interpretation of the data.  This necessitates asking the question of
whether these cross correlation techniques have unexplored
degeneracies, which might limit the accuracy at which planetary
properties can be discerned.  \citet{dek14} have studied the
conditions under which atmospheric composition and other planetary
properties can be recovered from high resolution transmission spectra
via cross correlation techniques.  However, the authors limited their
study to 1-D models (and 1D $\times$ 2 models where separate 1-D
profiles were used for the day side and the night side of the planet).
As we have demonstrated here, 3-D effects become discernible at high
spectral resolution, and it would therefore be unwise to ignore these
effects until their impact on spectral recovery methods have been
fully studied.  In certain cases --- especially for the most highly
irradiated planets in our study, as well as those on the CH$_4$ / CO
equilibrium boundary --- our models reveal a large enough discrepancy
between 1-D and 3-D models that the 3-D effects might alter the
interpretation of the spectra.

In the era of 30 to 40~m class telescopes, high resolution spectroscopy of
exoplanet atmospheres will be a valuable tool for exoplanet
characterization.  As we have shown, winds in exoplanet atmospheres can be diagnosed by high resolution spectroscopy, so long as the resolution of the spectrograph is at least of order $v/c$, where $v$ is the typical LOS wind speed.  For winds $\sim 1$ km~s$^{-1}$, this requires spectroscopy at a resolution of at least $10^5$.  In addition to spectroscopy of hot jupiters, it may
be possible to characterize habitable terrestrial planets \citep{sne13b, rod14} and even to extract planetary masses directly
from transmission spectra \citep{dew13}.  The use of ground-based
facilities with high resolution spectrographs, in conjunction with
space-based observations that are typically limited to much lower
spectral resolution, will therefore allow for detailed studies of
exoplanet atmospheres that are not accessible at present.  This
includes observations of Doppler shifted transmission spectra, which
will provide valuable constraints on the nature of the atmospheric
circulation in exoplanet atmospheres.

\acknowledgements 
KH acknowledges support from the Swiss National Science Foundation and
the Swiss-based MERAC Foundation.

\bibliography{ms}

\begin{deluxetable}{lcccc}
%\centering
\tablecaption{Relevant parameters for each of the models \label{tab:params}}
\tablewidth{0pt}
%{\scriptsize
%\begin{tabular}{lccc}
%\hline\hline
%\multicolumn{1}{c}{Model} & \multicolumn{1}{c}{${\cal F}_0$ (W m$^{-2}$)}  & \multicolumn{1}{c}
%{$\Omega_p$ (s$^{-1}$)} & \multicolumn{1}{c}{a (AU)}  \\
%\hline
\tablehead{
\colhead{Model} & \colhead{${\cal F}_0$ (W m$^{-2}$)} & \colhead{$T_{irrad}$ (K)} & \colhead{$\Omega_p$ (s$^{-1}$)} & \colhead{a (AU)}
}
\startdata
C & $2 \times 10^4$ & 771 & $1.3 \times 10^{-6}$ & 0.28 \\
C1 & $5 \times 10^4$ & 969 & $2.6 \times 10^{-6}$ & 0.17 \\
C2 & $7 \times 10^4$ & 1054 & $3.4 \times 10^{-6}$ & 0.15 \\
W & $2 \times 10^5$ & 1370 & $7.5 \times 10^{-6}$ & 0.09 \\
W1 & $5 \times 10^5$ &1723 &  $1.5 \times 10^{-5}$ & 0.06 \\
W2 & $7 \times 10^5$ & 1874 & $1.9 \times 10^{-5}$ & 0.05 \\
H & $2 \times 10^6$ & 2437 & $4.2 \times 10^{-5}$ & 0.03 \\
H1 & $5 \times 10^6$ & 3064 & $8.4 \times 10^{-5}$ & 0.02 \\
%\hline
%\end{tabular}}\\
\enddata
\end{deluxetable}

%\begin{figure}
%\begin{center}
%\includegraphics[scale=0.65]{t_p.eps}
%\end{center}
%\caption{}%Global mean T-P profiles for the models presented in this paper...  } \label{fig:t_p_avg}
%\end{figure}

\end{document}